\documentclass[aps,prl,reprint,showpacs,amsmath,amssymb,superscriptaddress]{revtex4-1}
\usepackage{amsthm,amsmath,amsfonts,amssymb,amsxtra,appendix,bookmark,dsfont,latexsym}

\makeatletter
\usepackage{hyperref}
\usepackage{delarray,color}
\usepackage{euscript}

\theoremstyle{definition}

\theoremstyle{remark}

\usepackage{color}

\newcommand{\bdm}{\begin{displaymath}}
\newcommand{\edm}{\end{displaymath}}
\newcommand{\bdn}{\begin{eqnarray}}
\newcommand{\edn}{\end{eqnarray}}
\newcommand{\bay}{\begin{array}{c}}
\newcommand{\eay}{\end{array}}
\newcommand{\ben}{\begin{enumerate}}
\newcommand{\een}{\end{enumerate}}
\newcommand{\beq}{\begin{equation}}
\newcommand{\eeq}{\end{equation}}

\newcommand{\R}{\mathbb{R}}

\newcommand{\C}{\mathbb{C}}

\newcommand{\E}{\mathcal{E}}

\newcommand{\one}{{\ensuremath {\mathds 1} }}

\newcommand{\curl}{\mathrm{curl} \,}

\newcommand{\cLau}{c _{\rm Lau}}

\newcommand{\PsiLau}{\Psi_{\rm Lau}}

\newcommand{\MFf}{\E ^{\rm MF}}
\newcommand{\MFe}{E ^{\rm MF}}
\newcommand{\rhoMF}{\varrho ^{\rm MF}}

\newcommand{\gH}{\mathfrak{H}}
\newcommand{\be}{\mathbf{e}}
\newcommand{\bp}{\mathbf{p}}
\newcommand{\bx}{\mathbf{x}}
\newcommand{\by}{\mathbf{y}}
\newcommand{\bA}{\mathbf{A}}

\newcommand{\bV}{\mathbf{V}}
\newcommand{\PsiQH}{\Psi ^{\rm qh}}
\newcommand{\cQH}{c _{\rm qh}}

\renewcommand{\Im}{\mathrm{Im}}
\newcommand{\Heff}{H^{\mathrm{eff}}}
\newcommand{\Hplas}{H^{\mathrm{plas}}}
\newcommand{\bAe}{\mathbf{A}_{\mathrm{r}}}
\newcommand{\bAa}{\mathbf{A}_{\mathrm{a}}}
\newcommand{\rhoQH}{\varrho_{\mathrm{qh}}}
\newcommand{\Zqh}{Z_{\mathrm{qh}}}
\newcommand{\Fqh}{F_{\mathrm{qh}}}
\newcommand{\Fqhf}{\mathcal{F}_{\mathrm{qh}}}
\newcommand{\muqh}{\mu_{\mathrm{qh}}}
\newcommand{\lMF}{\lambda ^{\rm MF}}

\newcommand{\ellB}{l_B}

\date{March, 2016}

\begin{document} 

\title{Emergence of fractional statistics for tracer particles in a Laughlin liquid}

\author{Douglas Lundholm}
\affiliation{KTH Royal Institute of Technology, Department of Mathematics, SE-100 44 Stockholm, Sweden}
\email{dogge@math.kth.se}

\author{Nicolas Rougerie}
\affiliation{Universit\'e Grenoble 1 \& CNRS,  LPMMC (UMR 5493), B.P. 166, F-38 042 Grenoble, France}
\email{nicolas.rougerie@lpmmc.cnrs.fr}

\begin{abstract}
We consider a thought experiment where two distinct species of 2D particles in a perpendicular magnetic field interact via repulsive potentials. 
If the magnetic field and the interactions are strong enough, one type of particles forms a Laughlin state and the other ones couple to Laughlin quasi-holes. 
We show that in this situation, the motion of the second type of particles is described by an effective Hamiltonian, corresponding to the magnetic gauge picture for non-interacting anyons. 
The argument is in accord with, but distinct from, the Berry phase calculation of Arovas-Schrieffer-Wilczek. It suggests possibilities to observe the influence of effective anyon statistics in fractional quantum Hall systems. 
\end{abstract}


\maketitle

%


The basic explanation for the fractional quantum Hall effect~\cite{Girvin-04,Goerbig-09,StoTsuGos-99,Laughlin-99,Jain-07} is the occurence of strongly correlated fluids as the ground states of 2D electron gases under strong perpendicular magnetic fields. 
The elementary excitations (quasi-particles) of these peculiar states of matter carry a fraction of an electron's charge, leading to the quantization of the Hall conductance in certain fractions of $e^2/h$. 
Even more fascinating is the possibility that these quasi-particles may have fractional statistics, i.e. a behavior under continuous exchanges that interpolates between that of bosons and that of fermions~\cite{Khare-05,Myrheim-99,Ouvry-07,Wilczek-90}. Recently, it has been proposed~\cite{Cooper-08,BloDalZwe-08,MorFed-07,RonRizDal-11,Viefers-08} that this physics could be emulated in ultra-cold atomic gases subjected to artificial magnetic fields.

The main evidence for the emergence of fractional statistics concerns the quasi-hole excitations of the Laughlin wave functions~\cite{Laughlin-83}, which occur when the filling factor $\nu$ of the 2D electron gas is the inverse of an odd integer. 
Based on a Berry phase calculation, Arovas, Schrieffer and Wilczek~\cite{AroSchWil-84} argued that a continuous, adiabatic, exchange of two such quasi-holes leads the electrons' wave function to pick up a phase factor $e^{-i\pi\nu}$. 
This suggests that if the quasi-holes are to be considered as genuine quantum particles, they should be treated as anyons with statistics parameter $-\nu$.

While this constitutes a powerful argument, a more direct derivation of the emergence of fractional statistics seems desirable (see e.g. the discussion in~\cite{Forte-91}). 
It is indeed not entirely obvious that one should identify the change in the phase of the electrons' wave function with the statistics parameter attributed to the quasi-holes. 
Furthermore, the most striking consequences of quantum statistics --- 
the presence or absence of an exclusion principle (cf. \cite{LunSol-13b}), 
of condensation or a degeneracy pressure ---
are statistical mechanics effects that cannot be observed by measuring Berry phases.  

It has recently been proposed~\cite{ZhaSreGemJai-14,CooSim-15} that the fractional statistics of quantum Hall quasi-particles could be probed by observing the behavior of test particles immersed in an atomic gas forming a quantum Hall droplet. It thus seems timely to revisit the emergence of fractional statistics in this context. 

In this letter we present a derivation of the anyonic nature of the Laughlin quasi-holes that does not appeal to the Berry phase concept, 
and suggest a mechanism by which the statistical mechanics influence of the anyon statistics could be directly ascertained. 
Our main result is to derive explicitly an effective, emergent, Hamiltonian for the test particles, see Equation~\eqref{eq:eff Hamil} below. We then propose an ansatz~\eqref{eq:GS trapped} for its ground state in a specific experimental regime, and compute the associated density~\eqref{eq:tracer profile}, a measurable quantity one could use to probe the emergence of fractional statistics.

\medskip

We consider, as in the proposals~\cite{ZhaSreGemJai-14,CooSim-15}, two different species of interacting 2D particles. The Hilbert space is 
\begin{equation}\label{eq:Hilbert}
\mathcal{H} ^{M+N} = L ^2 _{\rm{sym}} (\R ^{2M}) \otimes L ^2_{\rm{sym}} (\R ^{2N}),
\end{equation}
where $M$ is the number of particles of the first type and $N$ the number of particles of the second type. 
For definiteness we assume that the two types of particles be bosons, whence the imposed symmetry in the above Hilbert spaces. 
The following however applies to any choice of statistics for both types of particles.

We write the Hamiltonian for the full system as (spin is neglected)
$$
H_{M+N} = H_M \otimes \one + \one \otimes H_N + \sum_{j=1} ^M \sum_{k=1} ^N W_{12} (\bx_j-\by_k), 
$$
where
\begin{equation}\label{eq:start Hamil M}
H_M= \sum_{j=1} ^M \left( \frac{1}{2m}(\bp_{\bx_j} + e \bA (\bx_j)) ^2 \right) + \!\!\sum_{1\leq i<j\leq M}\! W_{11} (\bx_i-\bx_j)
\end{equation}
is the Hamiltonian for the first type of particles and 
\begin{equation}\label{eq:start Hamil N}
H_N= \sum_{k=1} ^N \left( \frac{1}{2}(\bp_{\by_k} + \bA (\by_k)) ^2  \right) + 
\sum_{1\leq i<j\leq N} W_{22} (\by_i-\by_j)
\end{equation}
that for the second type of particles. We shall denote $X_M= (\bx_1,\ldots,\bx_M)$ and $Y_N=(\by_1,\ldots,\by_N)$ the coordinates of the two types of particles and choose units so that $\hbar = c = 1$, and the mass and charge of the \emph{second} type of particles are respectively $1$ and $-1$. 
We keep the freedom that the first type of particles might have a different mass $m$ and a different charge 
$-e<0$. The first type of particles should be thought of as tracers immersed in a large sea of the second type. We shall accordingly use the terms ``tracer particles'' and ``bath particles'' in the sequel.

We have also introduced:
\begin{itemize}
 \item the usual momenta $\bp_{\bx_j} \!= - i \nabla_{\bx_j}$ and $\bp_{\by_k} \!= -i \nabla_{\by_k}$.
 \item a uniform magnetic field of strength $B >0$.
	Our convention is that it points downwards: 
 $$ 
	\bA (\bx) := -\frac{B}{2} \bx^{\perp}
	= -\frac{B}{2} (-x_2,x_1).
 $$
 \item intra-species interaction potentials, $W_{11}$ and $W_{22}$. 
 \item an inter-species interaction potential $W_{12}$.
\end{itemize}
The splitting between Landau levels of the one-body Hamiltonian appearing in $H_N$ is proportional to $B$, and we assume that it is large enough to force all bath particles to live in the lowest Landau level (LLL)
\begin{equation}\label{eq:intro LLL}
\gH = \left\{\psi (\bx) = f(z) e ^{ - B |z| ^2 / 4},\: f \mbox{ holomorphic } \right\}. 
\end{equation}
Note that the splitting between Landau levels for the tracer particles is rather proportional to $eB/m$ so if $m>e$ it is reasonable 
to allow that they occupy several Landau levels.

If in addition the bath particles' interaction potential $W_{22}$ is sufficiently repulsive, we are led to an ansatz of the form
\begin{equation}\label{eq:ansatz 1}
 \Psi (X_M,Y_N) = \PsiLau (z_1,\ldots,z_N) \Phi (X_M,Y_N)  
\end{equation}
for the joint wave function of the full system, where $\PsiLau$ is a Laughlin wave function:
$$ \PsiLau (z_1,\ldots,z_N) = \cLau \prod_{1\leq i<j \leq N} (z_i-z_j) ^{n}  \\ e ^{-  B \sum_{j=1} ^N |z_j| ^2 / 4} $$
with the coordinates $Z_N= (z_1,\ldots,z_N ) \in \C ^N$ identified with $\by_1,\ldots,\by_N \in \R ^2$. 
To stay within the allowed Hilbert space one must impose that $\Phi$ in~\eqref{eq:ansatz 1} be holomorphic in $\by_1,\ldots,\by_N$. 
We assume that tuning the integer $ n $ allows to completely cancel the interaction $W_{22}$ between particles of the second type. 
It is for example the case if $W_{22}$ has zero range and $ n $ is large enough~\cite{TruKiv-85,PapBer-01,RouSerYng-13b}. In the context of~\eqref{eq:Hilbert}, $ n $ should be even, $ n  = 2$ being the most relevant case.  For 2D electron gases, the second Hilbert space should be antisymmetric and $ n $ should be odd. The filling factor is then $ \nu = 1/ n $. 

Next we consider the situation where the inter-species interaction potential $W_{12}$ is strong enough to force the joint wave function of the system to vanish whenever particles of different species meet, i.e. $\Psi (\bx_j = \by_k) = 0$ for all $j,k$. 
Since $\Psi$ must stay within the lowest Landau level of the bath particles, this forces the form
$$ \Psi (X_M,Y_N) = \Phi (X_M,Y_N) \prod_{j=1} ^M \prod_{k=1} ^N (\zeta_j - z_k) \PsiLau (Z_N), $$
where we identify $\zeta_j \in \C$ with $\bx_j \in \R ^2$. 
We assume that choosing such an ansatz cancels the inter-species interaction, i.e. that the latter is sufficiently short range. 
Then, all terms of the total Hamiltonian acting on the bath particles are frozen and thus $\Phi$ depends only on $\bx_1,\ldots,\bx_M$. 
This leads us to our final ansatz
\begin{equation}\label{eq:ansatz 3}
 \Psi (X_M,Y_N) = \Phi (X_M) \cQH(X_M) \PsiQH (X_M,Y_N)
\end{equation}
where $\PsiQH$ describes a Laughlin state of the $N$ bath particles, coupled to $M$ quasi-holes at the locations of the tracer particles:
\begin{multline}\label{eq:QH}
\PsiQH (X_M,Y_N) 
 =  \prod_{j=1} ^M \prod_{k=1} ^N (\zeta_j - z_k)  \\ \prod_{1\leq i<j \leq N} (z_i-z_j) ^{ n }  e ^{- B \sum_{j=1} ^N |z_j| ^2 / 4}. 
\end{multline}
Here we choose $\cQH (X_M) >0 $ to enforce 
\begin{equation}\label{eq:norm QH}
\cQH (X_M) ^2 \int_{\R ^{2N}}  |\PsiQH (X_M,Y_N)| ^2 \,dY_N = 1 
\end{equation}
for any $X_M$. We thus ensure normalization of the full wave function $\Psi$ by demanding that 
$$ \int_{\R ^{2M}} |\Phi (X_M)| ^2 \,d X_M = 1.$$


We next argue that, for a wave function of the form~\eqref{eq:ansatz 3} and if $N\gg M$, $N\gg 1$, 
\begin{equation}\label{eq:main claim}
\left\langle \Psi, H_{M+N} \Psi \right\rangle \simeq \left\langle \Phi, \Heff_M  \Phi \right\rangle + \frac{B N}{2}
\end{equation}
for an effective Hamiltonian $\Heff_M$, so that the physics is completely reduced to the motion of the tracer particles. In this description, the interaction with the bath particles boils down to the emergence of effective magnetic fields in $\Heff_M$:
\begin{multline}\label{eq:eff Hamil}
\Heff_M= \sum_{j=1} ^M \frac{1}{2m}  \left( \bp_{\bx_j} + e \bA (\bx_j) + \bAe (\bx_j) + \bAa (\bx_j) \right)^2 
\\ + \sum_{1\leq i<j\leq M} W_{11} (\bx_i-\bx_j).
\end{multline}
Here the original potentials $\bA$ and $W_{11}$ are inherited from~\eqref{eq:start Hamil M}, while $\bAe$ and $\bAa$ emerge from the interaction with the bath particles. 
The subscripts stand for ``renormalizing'' and ``anyon'' vector potentials respectively. 
We have the expressions
\begin{equation}\label{eq:eff field ext}
 \bAe (\bx) = \frac{B}{2\pi  n } \int_{\R ^2} \frac{(\bx-\by)^{\perp}}{|\bx-\by| ^2} \one_{D(0,R)} (\by) \,d\by
\end{equation}
for some $R\propto \sqrt{N}$ to be defined below, and 
\begin{equation}\label{eq:eff field any}
 \bAa (\bx_j) = -\sum_{k=1,k\neq j} ^M \frac{B}{2\pi  n } \int_{\R ^2} \frac{(\bx_j-\by)^{\perp}}{|\bx_j-\by| ^2} \one_{D(\bx_k, \ellB)} (\by) \,d\by
\end{equation}
with the (conveniently scaled) magnetic length
$$ \ellB  = \sqrt{2/B}.$$
Everywhere in the paper we denote $D(\bx,R)$ the disk of center $\mathbf{x}$ and radius $R$, and $\one_{D(\bx,R)}$ the corresponding indicator function (equal to $1$ in the disk and $0$ outside).

Note that $\bAe$ corresponds to a constant magnetic field supported in a large disk, 
$$ \curl \bAe (\bx) =  \frac{B}{ n } \one_{D(0,R)} (\bx), $$
while $\bAa$ is generated by Aharonov-Bohm-like units of flux attached to the tracer particles' locations:
$$ \curl \bAa (\bx_j) = - \sum_{k=1,k\neq j} ^M \frac{B}{ n } \one_{D(\bx_k, \ellB)} (\bx_j).$$
In this convention, the wave function $\Phi$ still has the symmetry imposed at the outset, and the above thus describes anyons in the 
magnetic gauge picture~\cite{Wilczek-90,Myrheim-99,Khare-05,Ouvry-07}. 

We remark that the Aharonov-Bohm magnetic fluxes are naturally smeared over a certain length scale in the above (extended anyons model~\cite{Mashkevich-96,Trugenberger-92b,ChoLeeLee-92,LunRou-15}). 
However, the derivation strictly speaking requires that the disks 
in~\eqref{eq:eff field any} do not overlap, e.g. that the interaction potential $W_{11}$ contains a sufficiently extended hard core. 
Thus, by Newton's theorem one may replace  
\begin{equation}\label{eq:approx anyon}
 \bAa (\bx_j) = - \sum_{k=1,k\neq j} ^M \frac{1}{ n } \frac{(\bx_j-\bx_k)^{\perp}}{|\bx_j-\bx_k| ^2}.
\end{equation}
Finally, we shall see below that for $M\ll N$,  $R\propto \sqrt{N}$ is typically very large, so that one might want to further approximate
\begin{equation}\label{eq:approx ext}
 \bAe (\bx) =  \frac{B}{2 n } \bx ^{\perp}.
\end{equation}
The effect of this field is thus to reduce the value of the external one. This charge renormalization is due to the fractional charge associated with Laughlin quasi-holes.

\medskip

We turn to vindicating our claim~\eqref{eq:main claim}. 
Clearly, it is sufficient to show that, for any $j=1\ldots M$ we have essentially
\begin{equation}\label{eq:eff momentum} 
\bp_{\bx_j} \Psi \simeq \cQH \PsiQH \left( \bp_{\bx_j} + \bAe (\bx_j) + \bAa (\bx_j) \right) \Phi.
\end{equation}
Indeed, inserting this in the expression for the energy and recalling~\eqref{eq:norm QH} we may integrate first in the $Y_N$ variables to deduce~\eqref{eq:main claim}. 
The constant in~\eqref{eq:main claim} is just the LLL energy of the $N$ bath particles. 

We need to recall a few facts about the Laughlin and quasi-holes wave functions. Let us denote $\rhoQH$ the one-particle density of the quasi-holes ansatz,
\begin{multline}\label{eq:QH density}
\rhoQH (\by) = N |\cQH| ^2 \int_{\R ^{2(N-1)}} |\PsiQH (X_M,\by,\by_2,\ldots,\by_N)| ^2 \\
d\by_2\ldots d\by_N,
\end{multline}
which implicitly depends on $X_M$. Laughlin's plasma analogy (see the appendix for details) provides reasonable approximations for $\rhoQH$ and $\cQH$ in terms of an auxiliary classical mean-field problem, whose density we denote $\rhoMF$. 
Explicitly, assuming $N\gg 1$ and $N\gg M$, we approximate 
\begin{equation}\label{eq:MF density}
N ^{-1}\rhoQH \simeq \rhoMF = \frac{B}{2\pi  n  N} \left( \one_{D(0,R)} - \sum_{j = 1} ^M \one_{D(\bx_j, \ellB)} \right) 
\end{equation}
with $R ^2  = 2 ( n  N  +M)/B.$
We also assume that the disks 
$D(\bx_j, \ellB)$ in~\eqref{eq:MF density} do not overlap, e.g. because the tracer particles feel a hard core preventing them from getting too close to one another. 
One might still proceed without these assumptions, but we shall stick to the simplest case in this letter. 
Applying the Feynman-Hellmann principle to the effective plasma also leads to a useful approximation for the derivatives of $\cQH$ with respect to the location of the tracer particles:
\begin{equation}\label{eq:deriv cqh}
\frac{\nabla_{\bx_j} \cQH}{\cQH} \simeq N \int_{\R^2} \frac{\by-\bx_j}{|\by-\bx_j| ^2 } \rhoMF (\by) \,d\by.
\end{equation}

With these approximations at hand we can proceed to the derivation of~\eqref{eq:eff momentum}
where, from now on and without loss of generality, we take $j=1$. 
Clearly, 
$$
\nabla_{\bx_1} \Psi = (\nabla_{\bx_1} \Phi) \cQH \PsiQH + \Phi \left( \cQH \nabla_{\bx_1} \PsiQH + \PsiQH \nabla_{\bx_1} \cQH \right).
$$
A straightforward calculation shows that 
\begin{align*} 
\noalign{ $\nabla_{\bx_1} \PsiQH = \mathbf{V} (\bx_1) \PsiQH$, }
\mathbf{V} (\bx_1) 
&= \left(\sum_{k=1} ^N \frac{1}{\zeta_1-z_k} \right) \left( \be_1 + i \be_2 \right)\\
&= \sum_{k=1} ^N \frac{\bx_1-\by_k }{|\bx_1-\by_k|^2} + i\sum_{k=1} ^N \frac{(\bx_1-\by_k)^{\perp}}{|\bx_1-\by_k|^2}\\
&= \int_{\R^2} \left( \frac{\bx_1-\by}{|\bx_1-\by|^2} + i \frac{(\bx_1-\by)^{\perp}}{|\bx_1-\by|^2}\right)\!\left( \sum_{k=1}^N \delta_{\by=\by_k}\right)\!d\by
\end{align*}
where $\be_1$ and $\be_2$ are the basis vectors in $\R ^2$. 
In the state $\Psi$, the bath particles $\by_1,\ldots,\by_N$ are distributed according to the density $\rhoQH$, so that we may safely approximate
$$
\sum_{k=1} ^N \delta_{\by=\by_k} \simeq  \rhoQH (\by) \simeq N \rhoMF (\by)
$$
for the purpose of computing~\eqref{eq:main claim}. 
Inserting in the above, recalling~\eqref{eq:deriv cqh}, we observe that the real part of $\mathbf{V}$ cancels with $\nabla_{\bx_1} \cQH$, leading to 
$$ 
\bp_{\bx_1} \Psi = \cQH \PsiQH \left( \bp_{\bx_1} + \Im \mathbf{V} (\bx_1) \right) \Phi  
$$
where 
$$
\Im \mathbf{V} (\bx_1) \simeq N \int_{\R^2}  \frac{(\bx_1-\by) ^{\perp}}{|\bx_1-\by|^2} \rhoMF (\by) \,d\by.
$$
Inserting the expression~\eqref{eq:MF density} of $\rhoMF$ yields
\begin{align*}
\Im \bV (\bx_1) &= \frac{B}{2\pi n }\int_{\R^2}  \frac{(\bx_1-\by)^{\perp}}{|\bx_1-\by|^2} \one_{D(0,R)}(\by) \,d\by\nonumber\\
&-\frac{B}{2\pi n }\int_{\R^2} \frac{(\bx_1-\by)^{\perp}}{|\bx_1-\by|^2} \one_{D(\bx_1, \ellB)}(\by) \,d\by\nonumber\\
&-\frac{B}{2\pi n }\sum_{j=2}^M \int_{\R^2} \frac{(\bx_1-\by) ^{\perp}}{|\bx_1-\by|^2} \one_{D(\bx_j, \ellB)}(\by) \,d\by\nonumber.
\end{align*}
The second term is clearly $0$, while the first and third terms give the contributions of~\eqref{eq:eff field ext} and~\eqref{eq:eff field any} respectively. 
This closes the argument establishing~\eqref{eq:eff momentum}, and~\eqref{eq:main claim}
follows because of~\eqref{eq:norm QH}.

\medskip

We now discuss possible measurable consequences of the above derivation. 
The effective Hamiltonian~\eqref{eq:eff Hamil} is notoriously hard to solve, even at the ground state level 
(see~\cite{Wilczek-90,Myrheim-99,Khare-05,Ouvry-07} for reviews).
One can however use certain known results for comparisons with experiments. 
For illustration, we shall consider a case where the comparison seems the most simple and direct. 
We have a cold atoms system in mind, with both types of particles being bosons held by a harmonic confinement
$$ V(r) = \frac{1}{2} \omega ^2 r ^2.$$
The (artificial) magnetic field can be imposed by rotating the trap or by more elaborate means~\cite{DalGerJuzOhb-11}.
The previous discussion is unchanged if the energy scale associated with the trap is smaller than those entering the derivation, thus the effective Hamiltonian for the tracer particles is~\eqref{eq:eff Hamil}, supplemented by $m$ times the trapping term. 
As previously discussed we use the expressions~\eqref{eq:approx anyon} and~\eqref{eq:approx ext} for the effective fields. 
If $eB/m$ is large compared to the trapping energy, it makes sense to project this Hamiltonian onto the LLL of the associated effective magnetic field, of strength 
$$eB^* = (e- n  ^{-1})B.$$ 
The associated free Hamiltonian (case $W_{11} = 0$) is then essentially exactly soluble (see~\cite{DasOuv-94,Ouvry-07} and references therein). 
In the case of a weak interaction potential $W_{11}$, it is reasonable to expect that the ground state is also dictated by the free Hamiltonian, namely
\begin{equation}\label{eq:GS trapped}
 \Phi (X_M) = c_{\Phi} \prod_{1\leq i < j \leq M} \left|\zeta_i - \zeta_j\right|^{1/ n } e^{-\frac{m\omega_t}{2}\sum_{j=1} ^M |\zeta_j| ^2}, 
\end{equation}
see~\cite[Equations (6) and (36) to (38)]{Ouvry-07}. We have denoted 
$$ \omega_t = \sqrt{\omega ^2 + \left(\frac{eB^*}{2m}\right)^2}$$
an effective trapping frequency and $c_{\Phi}$ a normalization constant. 
Note that since~\eqref{eq:GS trapped} vanishes at $\zeta_i = \zeta_j$, we expect that this ansatz is still reasonable in the presence of a repulsive potential $W_{11}$. 

As a possible probe of the situation just discussed, we note that if the state of the full system is given by~\eqref{eq:ansatz 3}-\eqref{eq:norm QH}-\eqref{eq:GS trapped}, the one-particle density of the tracer particles can be approximated, for large $M$, as 
\begin{equation}\label{eq:tracer profile}
 \rho_{\rm tracer} \simeq \frac{m\omega_t  n  }{\pi} \one_{D(0,R')},
\end{equation}
where $R' = \sqrt{M/(m \omega_t  n )}$ is fixed by normalization. 
This follows from the same kind of considerations that lead to~\eqref{eq:MF density}, see the appendix for details. 
Both the distinctive flat profile and the (length and density) scales involved are signatures of the effective anyon Hamiltonian, and thus its emergence can be directly seen in a measurement of the density profile of the tracer particles. 
Indeed, both the gaussian profile one would get for free bosons in the LLL, and the Thomas-Fermi-like profile in the presence of weak pair interactions (see e.g. \cite{AftBlaDal-05,AftBlaNie-06a,AftBlaNie-06b}), differ markedly from~\eqref{eq:tracer profile}.

\medskip

\textbf{Conclusions}.
Assuming an ansatz of the form~\eqref{eq:ansatz 3} for the joint wave function of the system, we have demonstrated that the tracer particles feel an effective magnetic Hamiltonian~\eqref{eq:eff Hamil}. 
The latter contains long-range magnetic interactions whose form is identical to those appearing in the ``magnetic gauge picture'' description of anyons with statistics parameter $-1/ n $. 
One may thus (formally) extract from $\Phi$ a phase factor to gauge $\bAa$ away (singular gauge transformation). 
The effective Hamiltonian is then a usual magnetic Laplacian, with reduced magnetic field, but the effective wave function (formally) picks up a factor of $e^{-i\pi /  n }$ upon exchanging two particles, hence describes anyons. 

The assumptions we made are consistent with those of the usual Berry phase calculation~\cite{AroSchWil-84}. Namely, the energy scales associated to the effective Hamiltonian we derived should be smaller than the energy gap above the ground state ansatz~\eqref{eq:ansatz 3}.

%
%

The reasoning we proposed suggests a way to probe the anyon statistics, by a direct observation of the collective behavior of the tracer particles. 
If they are originally bosons, they will acquire some form of exclusion principle~\cite{DasOuv-94,Wu-94,LunSol-13a,LunSol-13b,LunSol-14} whose influence could be observed.


We have discussed a possible set-up where simple calculations show measurable effects of the emergent fractional statistics. 
The consideration of more general situations will demand further studies of the trapped anyon gas (see however~\cite{LunRou-15} and references therein for the discussion of approximate models). This remains a topic for future investigation, as does a more mathematically rigorous derivation of the phenomenon. 

As for possible generalizations, non-Abelian Quantum Hall states as considered in~\cite{ZhaSreJai-15} are not currently covered by our methods. However, if one is willing to take for granted, or argue for, appropriate replacements to~\eqref{eq:MF density}-\eqref{eq:deriv cqh}, the approach applies to the case where the bath particles form another Abelian Quantum Hall state, such as a composite fermions state as discussed  in~\cite{ZhaSreGemJai-14}. 

%

\bigskip 

\noindent\textbf{Acknowledgments:} This work is supported by the ANR (Project Mathostaq ANR-13-JS01-0005-01)
and the Swedish Research Council (grant no. 2013-4734). 
Many thanks to Thierry Champel, Michele Correggi, Stefano Forte, St\'ephane Ouvry, Gianluca Panati, Fran\c{c}ois D. Parmentier, Jan Philip Solovej and Jakob Yngvason for useful and stimulating discussions.

\appendix

\section{Appendix : The plasma analogy}\label{sec:app}

Here we provide some support for our claims about the approximation of $\rhoQH$ and $\cQH$ in Equations~\eqref{eq:MF density} and~\eqref{eq:deriv cqh} respectively. 
As mentioned previously, we argue these are reasonable provided $N\gg 1$, $N\gg M$ and 
\begin{align}\label{eq:balls}
D(\bx_j, \ellB) &\cap D(\bx_k, \ellB) = \varnothing \mbox{ for all } j\neq k \nonumber\\
D(\bx_j, \ellB) &\subset D(0,R) \mbox{ for all } j=1\ldots M.
\end{align}
Physically, these assumptions mean that the tracer particles should be thought of as a small number of impurities immersed in a large sea of the bath particles. That the disks 
around $\bx_j$ and $\bx_k$ do not overlap assumes a hard-core condition, that can be provided by the interaction potential $W_{11}$.

Laughlin's plasma analogy, originating in~\cite{Laughlin-83,Laughlin-87}, consists in writing 
$$ \cQH ^2 |\PsiQH| ^2 = \muqh = \frac{1}{\Zqh}\exp\left( -\Hplas (Y_N) \right)$$
as a Gibbs state for a classical electrostatic Hamiltonian $\Hplas$. In this convention, the partition function is $\Zqh$ and the temperature is $1$. Explicitly we have 
\begin{multline}\label{eq:Hplasma}
\Hplas (Y_N)= \sum_{j=1} ^N \left( \frac{B}{2}|\by_j| ^2 - 2 \sum_{k=1}^M \log |\by_j-\bx_k| \right) 
\\ - 2  n  \sum_{1\leq i < j \leq N} \log |\by_i - \by_j|, 
\end{multline}
which can be regarded as the Hamiltonian for $N$ 2D classical charged particles in a quadratic external potential, interacting via repulsive Coulomb forces. 
The quasi-holes/tracer particles appear in this representation as fixed repulsive point charges whose influence is felt by the bath particles.  

Thus, $\muqh$ is characterized as the probability measure minimizing the free energy
\begin{equation}\label{eq:free ener func}
\Fqhf [\mu] = \int_{\R ^{2N}} \Hplas (Y_N) \mu (Y_N) dY_N + \int_{\R ^{2N}} \mu \log \mu
\end{equation}
and the minimum free energy $\Fqh$ satisfies 
\begin{equation}\label{eq:free ener}
\Fqh = - \log \Zqh = 2 \log \cQH.
\end{equation}
Without further approximations, this rewriting is not particularly useful, but the point is that we may use the good scaling properties of $\PsiQH$, inherited from the fact that it is of the form polynomial $\times$ gaussian. 
Indeed, scaling length units by a factor $\sqrt{N}$ transforms the limit $N\to \infty$ for  the effective plasma problem into a mean-field/small temperature regime. 
That is, it gets mapped to yet another equilibrium statistical mechanics problem, but this time with a coupling constant $\propto N ^{-1}$ and an effective temperature $\propto N ^{-1}$. 

Since the minimization of~\eqref{eq:free ener func} can be mapped to a mean-field/small temperature regime by a simple change of scale, it is reasonable to use an ansatz of the form $\mu = \rho ^{\otimes N}$ and neglect the entropy to perform the minimization. 
We refer to~\cite{RouSerYng-13a,RouSerYng-13b,RouYng-14} where this approximation is rigorously justified for related models, and to~\cite{CaiLevWeiHan-82,CaiLev-86} for numerical confirmation. Making this manipulation leads to a classical mean-field energy functional 
\begin{multline}\label{eq:MF functional}
\MFf [\rho] = \int_{\R ^2 } \left( \frac{B}{2}|\bx| ^2 - 2 \sum_{j=1}^M \log |\bx-\bx_j| \right) \rho (\bx) \,d\bx \\
-  n  N \iint_{\R^2 \times \R^2} \rho (\bx) \log |\bx-\by| \rho (\by) \,d\bx d\by,
\end{multline}
and we should expect
$$ \rhoQH \underset{N\to\infty}{\simeq} N \rhoMF \mbox{ and } \Fqh \underset{N\to\infty}{\simeq} N \MFe$$
with $\MFe$ and $\rhoMF$ respectively the minimum and minimizer of~\eqref{eq:MF functional}. 
Furthermore, the Feynman-Hellmann principle tells us that the derivatives of $\MFe$ with respect to $\bx_j$ are given by integrating the derivative of the energy functional against the minimizer:
\begin{equation}\label{eq:deriv cqh 2}
\nabla_{\bx_j} \MFe = 2\int_{\R^2} \frac{\bx-\bx_j}{|\bx-\bx_j| ^2 } \rhoMF (\bx) \,d\bx.
\end{equation}
At this stage, combining with~\eqref{eq:free ener} we have thus justified (or, at least, explained) the approximations~\eqref{eq:MF density} and~\eqref{eq:deriv cqh}. 
What remains to be discussed is the explicit expression for $\rhoMF$ (right-hand side of \eqref{eq:MF density})
that we have used in the main text.
Note that the density profile \eqref{eq:tracer profile} for the tracer particles is obtained in the same way, by a simple change of units.

Recall that the expression is clearly physically sound: the $B|\bx|^2/2$ term in the energy functional~\eqref{eq:MF functional} corresponds to the potential generated by a constant background of charge density $-2B$ (in somewhat arbitrary units). 
Each tracer particle/quasi-hole corresponds to a fixed point with charge $4\pi$, and the bath particles have charge $4\pi  n  N$ in this representation. 
Each tracer particle thus screens the background charge in a disk 
of radius $ \ellB$ around it, and the bath particles distribute in order to screen the remaining background density.

For a more rigorous derivation, note that the minimizer $\rhoMF$ for~\eqref{eq:MF functional} must satisfy the variational inequality
\begin{equation}\label{eq:var eq MF}
\frac{B}{2}|\bx| ^2 - 2 \sum_{j=1}^M \log |\bx-\bx_j| - 2  n  N \rhoMF * \log|\, . \, | \geq \lMF
\end{equation}                                                                         
with equality on the support of $\rhoMF$. Here,
$$ \lMF = \MFe -  n  N \iint_{\R^2 \times \R^2} \rhoMF (\bx) \log |\bx-\by| \rhoMF (\by) \,d\bx d\by$$
is a Lagrange multiplier. A useful result is the following (see~\cite[Theorem 1.2]{ChaGozZit-13}): let $\rho$ be a probability measure that satisfies, for some $\lambda \in \R$,
\begin{equation*}
\frac{B}{2} |\bx|^2 - 2 \sum_{j=1}^M \log |\bx-\bx_j| - 2  n  N \rho * \log|\, . \, | \leq \lambda \ \mbox{($\geq \lambda$)}
\end{equation*}
where $\rho>0$ (respectively $\rho=0$).
Then it must be that
$\lambda = \lMF$ and $\rho = \rhoMF$.
It is easy to see that the right-hand side of~\eqref{eq:MF density} satisfies the above, provided~\eqref{eq:balls} holds. 
Indeed, the potential $\rho * \log |\, . \,|$ can in this case be easily computed using Newton's theorem~\cite[Theorem~9.7]{LieLos-01}.



%

\end{document}